\documentclass[12pt]{iopart}
\expandafter\let\csname equation*\endcsname\relax
  \expandafter\let\csname endequation*\endcsname\relax
\usepackage{braket,colortbl,cleveref,amsthm,amsmath,amssymb,txfonts}
\usepackage[unicode=true,pdfusetitle, bookmarks=true,bookmarksnumbered=false,bookmarksopen=false, breaklinks=false,pdfborder={0 0 0},backref=false,colorlinks=false] {hyperref}
\hypersetup{ colorlinks,linkcolor=myurlcolor,citecolor=myurlcolor,urlcolor=myurlcolor}
\definecolor{myurlcolor}{rgb}{0,0,0.7}
\usepackage{graphics,graphicx}
\usepackage{color}
\usepackage{mathpazo}
\usepackage[format = hang,labelfont = bf,up, textfont = normal , up, justification =centerlast, singlelinecheck = true]{caption}
\usepackage{subfigure}
\usepackage{latexsym,verbatim,revsymb,color}

\usepackage{hyperref}
\usepackage[svgnames]{pstricks}
\usepackage{pst-solides3d}  
\theoremstyle{plain}

\def\bea{\begin{eqnarray}}
\def\eea{\end{eqnarray}}
\def\ba{\begin{array}}
\def\ea{\end{array}}

\def\ket{\rangle}
\def\bra{\langle}
\def\beq{\begin{eqnarray}}
\def\eeq{\end{eqnarray}}

\begin{document}

\title{Coherence makes quantum systems 'magical'}

\author{Chiranjib Mukhopadhyay,  Sk Sazim, and Arun Kumar Pati} 
\address{Quantum Information and Computation Group, Harish Chandra Research Institute, Homi Bhabha National Institute, Allahabad 211019, India}

\begin{abstract} 
\noindent Two primary facets of quantum technological advancement  that hold great promise are quantum communication and quantum computation. For quantum communication, the canonical resource is entanglement. For quantum gate implementation, the resource is `magic' in an auxiliary system. It has already been shown that quantum coherence is the fundamental resource for the creation of entanglement. We argue on the similar spirit that quantum coherence is  the fundamental resource when it comes to the creation of 'magic'. This unifies the two strands of modern development in quantum technology under the common underpinning of existence of quantum superposition, quantified by the coherence in quantum theory.  We also obtain a coherence monotone from the discrete Wigner distribution  We further study the interplay between quantum coherence and 'magic' in a qutrit system and that between quantum entanglement and 'magic' in a qutrit-qubit setting.

 
\end{abstract}

\maketitle

\section{Introduction}

What makes quantum technologies so much more powerful than their classical counterparts ? The answer must ultimately lie in the postulates of quantum theory. Especially the linear superposition principle, implying the existence of quantum coherence, can be intuitively thought of as the driving agent behind any quantum advantage. The recent quantification of superposition through the resource theory of quantum coherence \cite{baumgratz,winter, theurer, noc} has allowed us to formalize this intuition in a more rigorous way. Quantum entanglement, the basic resource behind quantum communication schemes like dense coding \cite{dense}, teleportation \cite{teleport} or remote state preparation \cite{rsp}, has been connected with quantum coherence  \cite{uttament}. However, quantum technologies are not limited to communication schemes. One of the principle scientific developments in last few decades has been the emergence of the theory of quantum computation as a more powerful  alternative to the paradigm of classical computation. Resources like quantum entanglement and quantum coherence have at various points been shown to lead to quantum advantage vis-a-vis classical computers.  For example, to implement Shor's algorithm, one needs a large amount of entanglement \cite{akp1}, whereas, to implement Grover's algorithm, one needs a small amount of entanglement \cite{akp2,akp22}. Similarly, it has recently been demonstrated that in order to implement the Deutsch-Jozsa algorithm \cite{dj}, one requires coherence as a resource \cite{hillary}. Quantum coherence has also been related to the success probability of the Grover search algorithm \cite{fan, namit}.  However, for a quantum computer to work, we must ultimately be able to implement quantum logic gates. Utilising auxiliary quantum states which are outside the stabilizer polytope and in a so called `magic' state can, in the context of stabilizer computation, enable the implementation of gates which are not classically simulable, e.g., the T-gate. A resource theory for such magic states was recently proposed \cite{njp_magic} and is a topic of active interest \cite{howard, goursanders}. In this article, we ask the following question - how does quantum coherence relate to 'magic' in quantum states. Firstly, starting with a brief recapitulation of resource theories of coherence and stabilizer states, we show, using contractive distance based monotones, that the 'magic' generated in a quantum state through incoherent operations \cite{baumgratz} is upper bounded by the amount of coherence initially in the state. Subsequently, we prove that the maximum amount of 'magic' generated through such incoherent operations can, by itself, be shown to be a coherence monotone. We further obtain a full coherence monotone based on the  discrete Wigner function representation \cite{wootters, chaturvedi, gross} of quantum states, the latter being useful for providing a calculable measure of 'magic'.  Next, we propose the counterparts to various types of incoherent operations in the resource theory of magic states.  We subsequently move on to revealing the link between 'magic' and other quantum resources like quantum coherence and entanglement, in small quantum systems. Finally, we outline some possible directions of future work.

\section{Resource theories of quantum coherence and magic}

In this section, we briefly remind the reader about the resource theories of quantum coherence, as laid down in Ref. \cite{baumgratz} and 'magic', as introduced in Ref. \cite{njp_magic}.  

\subsection{Resource theory of coherence}

The resource theory of coherence seeks to quantify the amount of superposition in quantum states with respect to a fixed basis, say $\lbrace |i\ket \rbrace$. Thus, the set $\mathcal{I}$ of free states, hence called $incoherent$ $states$, consists of purely classical mixtures of eigenkets, i.e., states of the form $\sigma = \sum_{i} c_i |i\ket\bra i|$. The free operations, dubbed $incoherent$ $operations$, are defined as being those CPTP operations whose every Kraus element maps an incoherent state to another incoherent state. One can easily see that this  is a stronger condition than merely requiring the CPTP operations to map every incoherent state to another incoherent state. The necessary conditions that any monotone $C[\rho]$ must satisfy under incoherent operation are thus given by

\begin{enumerate}
\item If $\sigma \in \mathcal{I}$, then $C[\sigma] = 0$. Otherwise, $C > 0$.

\item For any incoherent operation $\Lambda_{IC}$ and any state $\rho$, $C[\Lambda_{IC} (\rho)] \leq  C [\rho]$.

\item If $\lbrace K_i \rbrace$ are Kraus operators corresponding to any incoherent operation $\Lambda_{IC}$ such that $\sigma_{i} = \frac{K_{i} \rho K_{i}^\dagger}{\tr[K_{i} \rho K_{i}^\dagger]} = \frac{K_{i} \rho K_{i}^\dagger}{p_i} $, then the coherence should not increase under selective measurement, i.e. $\sum_i p_i C[\sigma_i] \leq C [\rho]$.
\end{enumerate}

Many examples of such coherence monotones have been found in literature \cite{baumgratz, ma,roc1,roc2, rana, rastegin,max, colloquium}.

\begin{figure}
\centering
\includegraphics[scale=0.7]{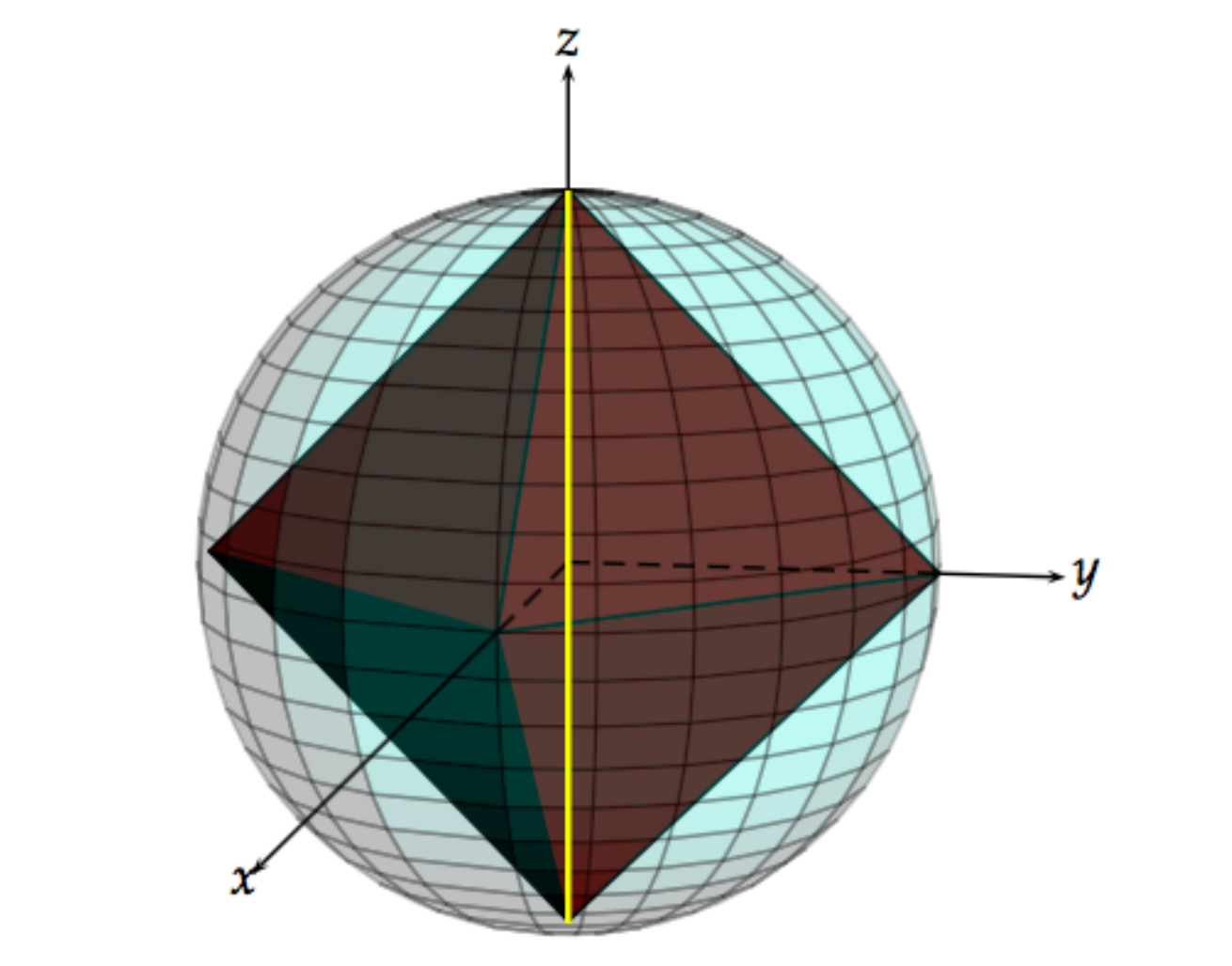}
\caption{(Color online) Pictorial representation of free states in resource theories of 'magic' and coherence in the qubit case. The stabilizer polytope is an octahedron within the Bloch sphere. Any qubit state represented by a point outside the octahedron is a magic state. All incoherent states in the computational basis lie on the yellow line.}
\end{figure}

\subsection{Resource theory of magic states}

The main aim behind the resource theory of magic states, also known as the resource theory of stabilizer computation, is to quantitatively characterize the extent to which a quantum system, acting as an auxiliary, can help in implementing classically non-simulable gates. The pure free states in this resource theory are the ones reachable via Clifford unitaries acting on any member of the computational basis, say $|0\ket$. The total family of free states, denoted as $stabilizer$ $states$ $\mathcal{S}$, consists of the convex hull formed by the pure free states.  The free operations consist of Clifford unitaries, measurement in the computational basis, composition with other stabilizer states and partial trace, as well as these operations conditioned on measurement results.  Magic monotones are relatively less studied until now, although some monotones have been found ranging from distance based monotones \cite{njp_magic} to robustness type monotones \cite{bartosz} to monotones inspired from the Wigner function representation of states in discrete phase space \cite{njp_magic}.
\\

As a  succinct reminder, the table below summarizes the primary features for both the resource theories.
\begin{center}
\begin{tabular}{ | m{5em} | m{5cm}| m{7cm} | } 
\hline
\textbf{Resource theory }&  \textbf{Free states }& \textbf{Free operation} \\ 
\hline
Coherence & diagonal incoherent states $\sigma = \sum_{i} c_{i} |i\ket\bra i| $ in the basis $\lbrace|i\ket \rbrace$ & Incoherent operation \begin{itemize}
\item  If $\sigma \in \mathcal{I}$, $\Lambda_{IC}[\sigma] \in \mathcal{I}$
\item If $\sigma \in \mathcal{I}$ then for each Kraus channel $\lbrace K_i \rbrace$ corresponding to incoherent operation, $K_i \sigma K_i^{\dagger} \in \mathcal{I}$
\end{itemize} \\ 
\hline
Magic & States inside polytope accessible via Clifford unitary rotation of computational basis & Stabilizer operation \begin{itemize}
\item Clifford unitary
\item Measurement in computational basis
\item Partial trace
\item Composition with other stabilizer states
\end{itemize} \\ 
\hline

\end{tabular}
\end{center}

\section{Linking resource theories of coherence and 'magic' }

\subsection{Coherence Quantifiers through Magic Monotones}
The resource theories of coherence and magic states, as reviewed above, seem quite disjoint. But are they really so ? This is the question we seek to address. Specifically, in this section, we demonstrate how the presence or lack of quantum coherence in systems constrains the amount of 'magic' in the system. In doing so, we reveal that quantum coherence can be quantified by the maximum amount of 'magic' generated through incoherent operation on arbitrary quantum states. In subsequent work, unless otherwise stated, the basis with respect to which quantum coherence is defined, is the computational basis and the pure stabilizer states are the ones obtainable through Clifford unitary rotation of one of the basis elements, say $|0\ket$, of the computational basis.  We now state our first result -

\textbf{Result 1.} \emph{For any distance based coherence quantifier $C_D$ and corresponding magic quantifier $M_D$, the amount of magic generated through incoherent operations $\Lambda_{IC}$ on a quantum state is upper bounded by the amount of coherence originally present in that state.}
\beq
M_{D} \left[ \Lambda_{IC} (\rho) \right] \leq C_{D} \left[\rho\right]
\label{cantincrease}
\eeq

\emph{Proof-} lhs equals $\min_{\mu \in \mathcal{S}} D[\Lambda_{IC} (\rho), \mu] \leq \min_{\sigma \in \mathcal{I}} D[\Lambda_{IC} 
(\rho) , \sigma ] = C_{D} [\Lambda_{IC}(\rho)] \leq C_{D} [\rho] $ 
where we used the fact that any incoherent state in the computational basis is a stabilizer state.                       \qed

We now propose the following set of coherence monotones corresponding to every distance based magic monotone

\beq 
C_{M} \left[\rho \right]= \sup_{\Lambda_{IC}} M\left[ \Lambda_{IC} (\rho) \right]
\label{bhou}
\eeq 
and prove the monotonicity conditions below. It is trivial to see that any incoherent state with respect 
to the adequately chosen basis is a stabilizer state. Monotonicity under CPTP maps is guaranteed for any 
contractive distance based measure. Therefore we only present the proof for strong monotonicity under 
selective measurements. The proof is identical in spirit to the one presented in \cite{uttament} for entanglement.

\textbf{Result 2 (Strong Monotonicity).} \emph{If $\sigma_{i} = \frac{K_{i} \rho K_{i}^\dagger}{\tr[K_{i} \rho K_{i}^\dagger]}$ 
and $p_{i} = \tr[K_{i} \rho K_{i}^\dagger]$ where $\lbrace K_{i}\rbrace$ are the Kraus operators 
corresponding to some incoherent operation, then }
\beq \sum_{i} p_{i} C_{M} [\sigma_{i}] \leq C_{M} [\rho]. \eeq

\emph{Proof-} Let us assume that the condition above is false. Then there will exist at least one set of incoherent 
operations $\lbrace \Lambda_{i} \rbrace$ for which 
\beq \sum_{i} p_{i} M [\Lambda_{i}\sigma_{i}] > C_{M} [\rho]. \eeq
Please note that each $\Lambda_i$ here is individually an incoherent operation and should not be confused as being merely one Kraus element of a  quantum operation.
Now, since magic monotones are non-increasing on average under measurements in the computational basis , 
\bea M \left[\sum_{i} p_{i} \rho_{i} \otimes |i\ket \bra i|\right] \geq \sum_{i} p_{i} M[\rho_{i}] \\
\Rightarrow M\left[ \sum_{i} p_{i} \Lambda_{i}\sigma_{i} \otimes |i\ket\bra i|  \right] > C_{M} [\rho] \\
\Rightarrow M \left[ \sum_{i} \Lambda_{i} \left(K_{i} \rho K_{i}^\dagger\right) \otimes |i\ket\bra i| \right]  > C_{M} [\rho].\eea
Where the last step  follows from the definition of $\sigma_i = K_i \rho K_i^{\dagger}/ p_i$. 
Now, one can write a bipartite incoherent operation $\tilde{\Lambda}$ such that the Kraus operators for  $\tilde{\Lambda}$ are written as $M_{ij} = L_{ij} (K_i) \otimes U_i $, where $L_{ij}$ are the Kraus operators corresponding to the incoherent operation $\Lambda_i$ and $U_i$ is the incoherent unitary $\sum_{j} | \text{mod} (i+j, \text{dim (ancilla)}) \ket \bra j|$. For this operation, the LHS = \beq M\left[\tilde{\Lambda} \left( \rho \otimes |0\ket\bra 0|  \right) \right]> C_{M}[\rho] 
=  C_{M} [\rho \otimes |0\ket \bra 0|]. \eeq This is in contradiction with the result \eqref{cantincrease} proved earlier , thus completing the proof. \qed
\subsection{Coherence monotone inspired from another magic monotone}

Historically, phase space methods in quantum optics and continuous variable quantum information theory, have been very successful. In particular, the phase space (quasi)-probability distributions like Wigner distribution, Sudarshan-Glauber $p$-distribution or Husimi $q-$distribution are extremely helpful in characterizing optical states \cite{scully}. Of these, the Wigner distribution is particulatly notable for the fact that it additionally reproduces the correct marginal probability distributions. Since the introduction of phase space distributions are so successful for CV systems, it was inevitable that attempts to create analogues of such distributions for qudit states were to be made. There exist many such proposed constructions in the literature, \cite{disc1,disc2, disc3}, of which we shall make use of the construction  of discrete Wigner function by Wootters \cite{disc1}.

In the preceding subsection, we showed how to construct coherence monotones from distance based magic monotones. In most cases, these monotones are extremely hard to exactly calculate. There is however, a computable monotone, called sum negativity, already in the literature \cite{njp_magic} in terms of the negativity of the discrete Wigner function representation of a state. We show that the discrete Wigner function representation can also give rise to a coherence monotone. Something similar was possibly attempted in Ref. \cite{chinese_wigner}. However, their paper (sans abstract) was written in Chinese, hence not accessible to us or anybody not familiar with the language. Therefore, we furnish a complete proof for the sake of completeness. 

For finite Hilbert space dimension $d$, the expression for characteristic function associated with each point $(p,q)$ on the $d \times d $ phase grid is given by \beq A_{(p,q)} =  D_{p,q} A_{0} D_{p,q}^\dagger, \eeq \vspace{0.1in}where 
$D_{p,q} = \omega^{- 2^{-1} pq} Z^{p} X^{q}$ and $A_{0} = \frac{1}{d}\sum_{p,q=0}^{d-1} D_{p,q}$. $X$ and $Z$ are the well known Shift and Boost operators respectively, and $\omega = e^{2 \pi i / d}$ is the $d$-th root of unity and $2^{-1}$ is shorthand for $\frac{d+1}{2} \mod (d)$.  The Wigner function  of a quantum state represented by the density matrix $\rho$, at a phase space point $(p,q)$, is given by $W_{(p,q)} = \tr (\rho A_{(p,q)})$. Sum of Wigner functions along a line $W_{q} = \sum_{p} W_{(p,q)}$is always positive semidefinite. Now let us propose the following candidate for a coherence monotone \beq C_{w} [\rho] = \min_{ \sigma \in \mathcal{I}, \lambda \geq 0} ||\vec{K}_{\rho} - \lambda \vec{K}_{\sigma}  || .\eeq Here $\vec{K}_{\rho}$ is a probability vector whose elements are the sums of Wigner functions $(W_1 (\rho), W_2 (\rho) , ....)$ along parallel lines in the phase grid and $|| P - Q ||$ is the statistical distance between probability distributions $P$ and $Q$.

Clearly, $C_{w}[\rho]$ vanishes for incoherent states. Moreover, from the monotonicity of trace distance under CPTP maps, $C_{w}$ is monotonically decreasing under any CPTP map. The remaining, i.e., strong monotonicity and convexity conditions have been shown in literature \cite{tong} to be equivalent to the equality condition $C[p \rho_1 \oplus (1-p) \rho_2] = p C[\rho_1] + (1-p) C[\rho_2]$. The LHS of the above condition now reads as $
C_{w}[p \rho_1 \oplus (1-p) \rho_2] = \min_{ \sigma_1, \sigma_2 \in \mathcal{I}, \lambda_1, \lambda_2 \geq 0} ||\vec{K}_{p \rho_1 \oplus (1-p) \rho_2} -  \vec{K}_{\lambda_1 \sigma_1 \oplus \lambda_2 \sigma_2 }  || =  \min_{ \sigma_1, \sigma_2 \in \mathcal{I}, \lambda_1, \lambda_2 \geq 0} ||p \vec{K}_{\rho_1 } +  (1-p) \vec{K}_{\rho_{2}} -  \lambda_1 \vec{K}_{\sigma_{1} } - \lambda_2 \vec{K}_{\sigma_2}   || =  \min_{ \sigma_1, \sigma_2 \in \mathcal{I}, \lambda_1, \lambda_2 \geq 0} ||p \vec{K}_{\rho_{1} } -  \lambda_1 \vec{K}_{\sigma_1 } + (1-p) \vec{K}_{\rho_2}- \lambda_2  \vec{K}_{\sigma_2}   ||  = p \min_{ \sigma_1 \in \mathcal{I}, \lambda_1' \geq 0} ||\vec{K}_{\rho} - \lambda_1' \vec{K}_{\sigma_1}  || + (1-p) \min_{ \sigma_2 \in \mathcal{I}, \lambda_2' \geq 0} ||\vec{K}_{\rho_2} - \lambda_{2}' \vec{K}_{\sigma_2}  ||  = p C_{w}[\rho_1] + (1-p) C_{w} [\rho_2]$ where $\lambda_{1}' = \lambda_1 / p$ and $\lambda_{2}' = \lambda_2 / (1-p)$.  This completes the proof of the assertion that $C_{w}$ is a full coherence monotone. 

It is natural to ask the question - what does $C_w$ signify physically ? We do not have a clear-cut answer and invite the reader to ponder upon this question, while remarking that our construction of $C_w$ is somewhat similar to the construction of the 'modified' trace distance monotone of coherence \cite{tong}.  It would thus be interesting to compare the 'modified' trace distance monotone with $C_w$  for specific qudit systems.

\subsection{Hierarchy of Stabilizer Operations}

\begin{figure}
\centering
\includegraphics[scale=0.35]{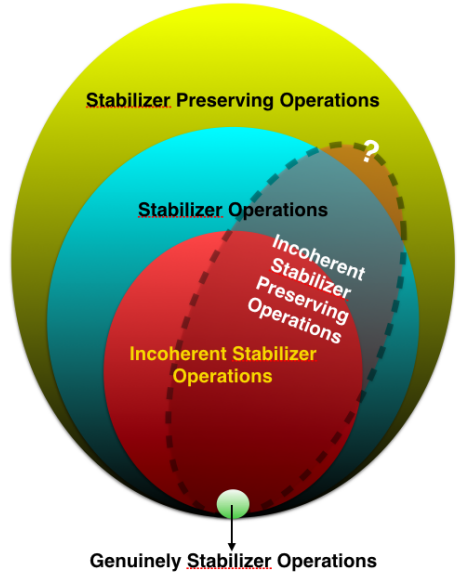}
\caption{(Color online) Hierarchy of various free operations in the resource theory of magic}
\label{fig_hierarchy}
\end{figure}
In analogy with the resource theories of quantum coherence or entanglement, we may formulate various generalizations and specializations of stabilizer operations. A  tentative hierarchy of such operations, roughly following the corresponding formulation for  incoherent operations in Ref. \cite{hier1,hier2} is depicted in Fig. \ref{fig_hierarchy}.

\noindent \emph{Genuinely Stabilizer Operations -} The most stringent of all the stabilizer operations must be the genuinely stabilizer operations (GSO) similar to genuinely incoherent operations introduced in \cite{streltsov} for which every stabilizer state is supposed to be a fixed point for the dynamics. In the following proposition- we prove that such an operation is impossible unless it 
is the trivial identity transformation.

\emph{Proposition - There is no non-trivial Genuinely Stabilizer Operation.} 

\emph{Proof-}Let us illustrate the proof for $d=2$.  Suppose there is such a CPTP operation $\Lambda$ which is a Genuinely Stabilizer Operation. This implies $\Lambda$ is a genuinely incoherent operation with respect to both the eigenbasis of $\sigma_z$ and $\sigma_x$. Thus the Kraus operators corresponding to this  operation are diagonal in both $z$ basis as well as $x$ basis, which holds true only for the trivial identity operation.   \qed.

\noindent \emph{Incoherent Stabilizer Operations -} Stabilizer operations can still generate quantumness in the form of quantum coherence. Thus, if we are to construct a resource theory encompassing both the stabilizer formalism and quantum superposition, it is relevant to consider incoherent stabilizer operations. In the stabilizer protocol, two operations stand out as potentially generators of quantum coherence. One being the Clifford unitary operation, the other being composition with different stabilizer states. The other operations, viz. measurement in computational basis or partial trace, can easily be shown to be incoherent operations as well. Thus, we write down the following subset of these two operations -

\begin{enumerate}
\item \emph{Incoherent Clifford Unitary} - Defined as those clifford unitaries which do not generate quantum coherence, these now represent permutations of computation basis vectors. For example, in the qubit case, the bit-flip gate $\sigma_x$ or the phase-gate $\begin{pmatrix}
1 & 0 \\
0 & i
\end{pmatrix}$ are incoherent Clifford unitaries, but the Hadamard gate H = $\frac{1}{2}\begin{pmatrix}
1 & 1 \\
1 & -1
\end{pmatrix}$ is not.

\item \emph{Composition with other incoherent states} Incoherent states are by definition, within the stabilizer polytope, and composition with other incoherent states keeps quantum coherence fixed \cite{udit}. Thus this represents a suitable incoherent stabilizer operation.

\end{enumerate}

Clearly every coherence monotone is a monotone under this formalism. 

\emph{Proposition - } \emph{The $\text{l}_{p}$-norm is a monotone  under incoherent stabilizer operations for every $p \geq 1 $ .}

\emph{Proof -}  Every stabilizer protocol $\Lambda$ on a state $\rho_{S}$ can be expressed as $\Lambda[\rho_S] = \text{Tr}_{A'} \left[ U_{SA} \left( \rho_S \otimes \sigma_A\right) U_{SA}^{\dagger} \right] $ where $U$ is a Clifford unitary and $\sigma$ is an ancilla stabilizer state.  According to the conditons above, we must restrict $\sigma$ to the set of incoherent stabilizer states and $U$ to the set of incoherent Clifford unitaries. The effect of incoherent unitaries is either  to permute the basis labellings for coherence. or to lend a phase term to the off-diagonal elements of the density matrix, the latter, for example, is illustrated by an incoherent unitary $\begin{pmatrix}
1 & 0 \\
0 & i
\end{pmatrix}.$
Both of these effects  can be easily verified to leave the $l_p$-norm invariant. The $l_p$-norm of a state  $\rho$  is given by \beq C_{l_p} [\rho] = \left(\sum_{i}\sum_{j , i \neq j} |\rho_{ij}|^p \right)^{1/p}. \eeq Now, for an incoherent ancilla state $\sigma = \sum_{k} q_k |k\ket\bra k|, $ the $l_p$ norm   $C_{l_p} [\rho \otimes \sigma] = (\sum_{k} q_{k}^p)^{1/p} C_{l_p} [\rho] \leq  C_{l_p} [\rho]$, where we have used H{\"o}lder's inequality. Similarly one can also check using the triangle inequality for the $p$-norm, that partial tracing doesn't increase  $C_{l_p}$. Thus, $l_{p}$-norm is indeed a monotone for every $p \geq 1$ under incoherent stabilizer protocols. \qed

Operationally, it is natural to wonder about the strength of stabilizer operations vis-a-vis the strength of incoherent stabilizer operations. A simple example would be to consider the Hadamard operation, which is a stabilizer unitary, but can generate coherence in the computational basis.

\noindent \emph{Incoherent Stabilizer Preserving Operations-} Continuing in the spirit of connecting the two resource theories, one can impose on the set of incoherent operations only the constraint that it does not generate any 'magic' from stabilizer states. For example, the phase rotation is an incoherent operation, which may easily be seen to create 'magic' starting from a stabilizer state. The mathematical characterizaton of incoherent stabilizer preserving operations is beyond the scope of the present work and we invite the reader to embark upon the same. An especially pressing concern would be to identify whether there exists any incoherent stabilizer preserving operation which is not an incoherent stabilizer operation, or even a stabilizer operation.  Tentatively, Fig. \ref{fig_hierarchy} reflects that such operations are not ruled out, however, we have not succeeded in finding explicit counterexamples or proofs either refuting or supporting this statement.

\noindent \emph{Stabilizer Preserving Operations -} This is the most general type of free operation in the resource theory of magic that one can envisage. One only imposes the constraint that no stabilzer state is mapped to a magic state. In fact, such operations were studied in detail in Ref. \cite{goursanders} and a family of monotones derived.

\section{Concrete results in small quantum systems }

In this section, we shift our focus towards linking 'magic' with other quantum resources in low dimensional systems. The smallest dimension for which we have a concrete computable expression for 'magic' is $d=3$, which is expressed  via the sum negativity of discrete Wigner functions. Let us now look at the interplay between quantum coherence and 'magic' in this scenario. Since signature of the connection between 'magic' and contextuality has already been revealed \cite{kcbs,contextuality_supplies_magic}, our method of relating 'magic' to other resource theories connects contextuality inter alia with these resources. The nascent resource theoretic formulation of contextuality \cite{resourcecontextuality, wiring} can shed further light on the results we obtain here.

\subsection{Explicit expression for 'magic' in the qutrit case}
We know from the discrete version of Hudson's theorem, that at least in the odd-dimensional case, a pure state with a positive discrete Wigner distribution, must be a stabilizer state. Accordingly, a measure of 'magic' named \emph{sum negativity}, embodying the negativity of the discrete Wigner function for a given state, was proposed and proved as a magic monotone for odd prime power dimensions under stabilizer operations \cite{njp_magic}. It is defined as the following - 

\emph{For any quantum state with discrete Wigner distribution} $W_{u}$, \emph{the sum negativity  is the sum of absolute values of the negative elements of the discrete Wigner (quasiprobability) distribution. }

For the qutrit case, after some algebra, we have the following discrete Wigner distribution corresponding to a qutrit density matrix $\rho$.

\bea 
W_{_{(1,1)}} =  \frac{1}{3}  \left( 2 \lambda_{3} + \rho_{11}  \right), 
W_{_{(1,2)}} = \frac{1}{3} \left( 2 \lambda_{2} + \rho_{22} \right),
W_{_{(1,3)}} = \frac{1}{3} \left( 2 \lambda_{1} + \rho_{33} \right),\nonumber \\
W_{_{(2,1)}} = \frac{1}{3} \left( - \lambda_{3} - \sqrt{3} \mu_{3} + \rho_{11} \right), W_{_{(2,2)}} = \frac{1}{3} \left( - \lambda_{2} + \sqrt{3} \mu_{2} + \rho_{22}\right),\nonumber \\
W_{_{(2,3)}} = \frac{1}{3} \left( -\lambda_{1} - \sqrt{3} \mu_{1} + \rho_{33} \right), W_{_{(3,1)}} = \frac{1}{3} \left( - \lambda_{3} + \sqrt{3} \mu_{3} + \rho_{11} \right),\nonumber \\
W_{_{(3,2)}} = \frac{1}{3} \left( - \lambda_{2} - \sqrt{3} \mu_{2} + \rho_{22}\right), W_{_{(3,3)}} = \frac{1}{3} \left( -\lambda_{1} + \sqrt{3} \mu_{1} + \rho_{33} \right).
\eea

Where density matrix elements  $\rho_{12} = \lambda_{1} + i \mu_{1}, \rho_{13} = \lambda_{2} + i \mu_{2}, \rho_{23} = \lambda_{3} + i \mu_{3}$, $\lambda_i , \mu_i \in \mathbb{R}$. 
Now the sum negativity $M_{SN} [\rho]$ is simply given by \beq  M_{SN}[\rho] = \sum_{\textbf{u}}|W_{\textbf{u}}| - 1 .\eeq Similar to the case of logarithmic negativity in the resource theory of entanglement, it is possible to come up with another monotone, which is the logarithm of the sum negativity. This measure was dubbed the $Mana$ \cite{njp_magic}.

\subsection{Effect of coherent and incoherent noise on magic states }

As with many other quantum resource theories, the maximally mixed state is a free state in the resource theory of magic states while the maximally resourceful state turns out to be a pure state. In the qutrit scenario, the maximally magical pure states come in two different varieties, viz. the Strange states \cite{njp_magic} and the Norrell states \cite{njp_magic}. The strange states are pure states which are  invariant under the symplectic component of the Clifford group and of the form $\frac{1}{\sqrt{2}} (0,1, \pm 1)^{T}$ and the corresponding permutations. Geometrically speaking, the Strange states are the pure states maximally distant from the \emph{faces} of the stabilizer polytope. The Norrell states, in contrast, are pure states maximally distant from the \emph{edges} of the stabilizer polytope. They are thus natural qutrit  generalizations to the $T$ and $H$ states for qubits respectively.

It may therefore be interesting to have an answer to the question that which  class  of states remain more magical under admixture of noise. However, noise can be either coherent or incoherent. As we demonstrate below, depending on the character of the noise, the relative robustness of two types of magical states may be of different nature.

\emph{Proposition - Strange states are more robust under mixture with maximally incoherent, i.e. white noise than Norrell states.}

\begin{figure}
\subfigure[White noise]{
    \includegraphics[width=0.45\textwidth, keepaspectratio]{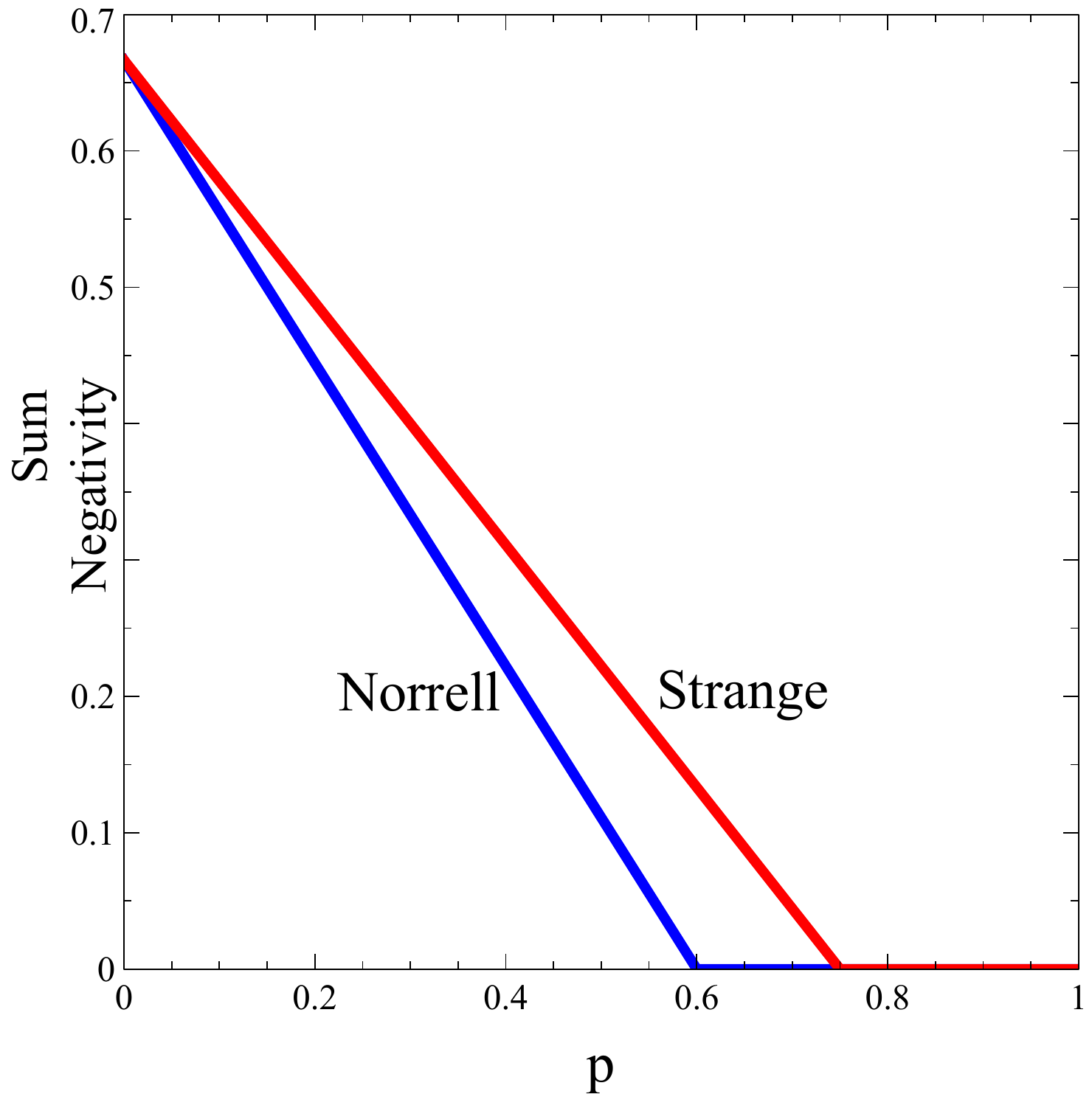}
    \label{white}
}
\subfigure[Coherent noise]{
    \includegraphics[width=0.45\textwidth, keepaspectratio]{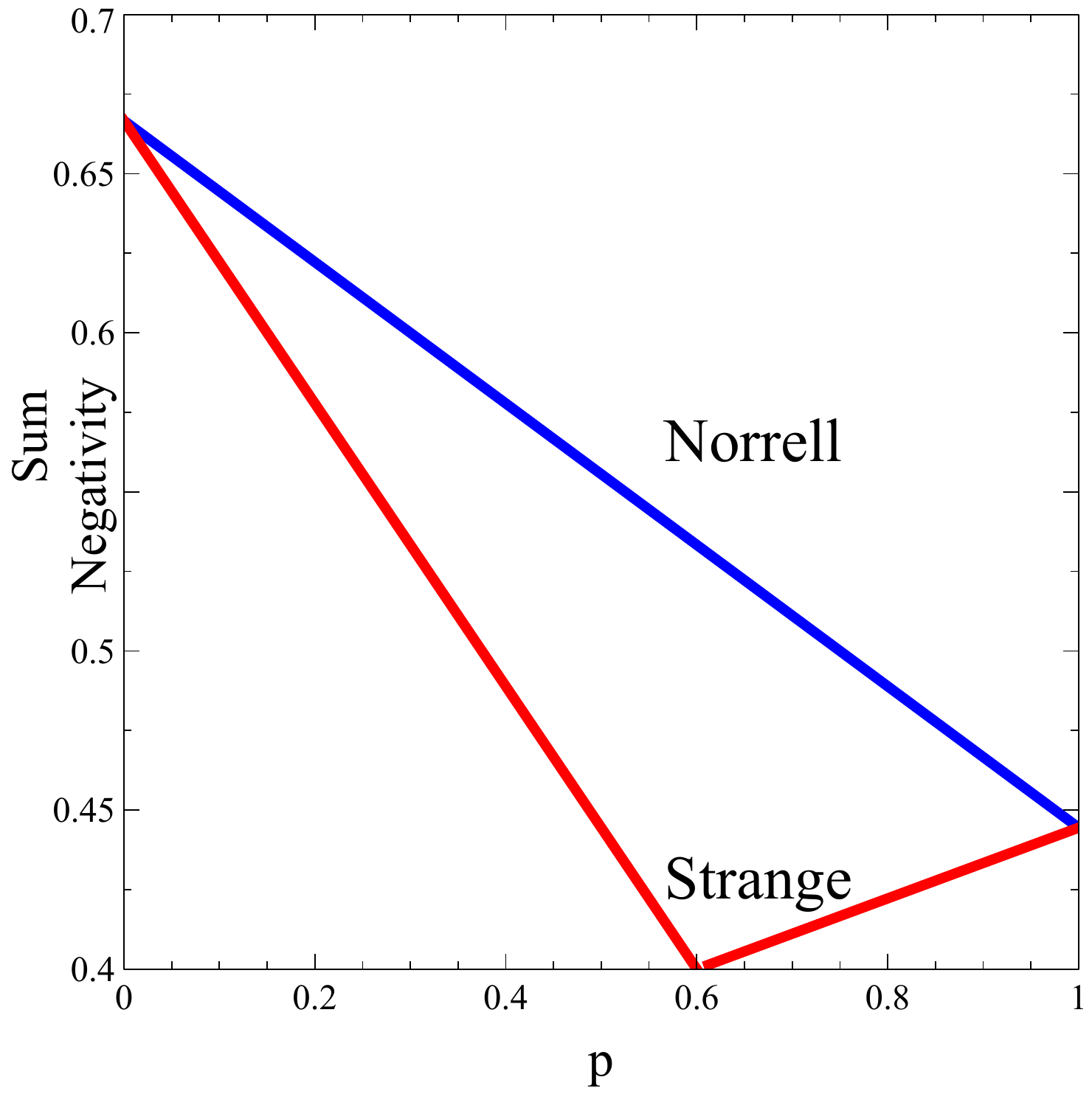}       
    \label{cohnoise}
}
\caption{(Color online) Response of two different maximally magical pure states, viz a strange state (red), and a Norrell state (blue), for admixture with maximally incoherent noise (left) and coherent noise (right). The sum negativity of the resulting states has been plotted with respect to the admixture parameter $p \in [0,1]$. }
\label{noise}
\end{figure}

\emph{Proof-}
Due to symmetry, it suffices to check for one strange state and one Norrell state, respectively. Let this strange state be $|\psi_S\ket = \frac{1}{\sqrt{2}} ( 0,1 , -1)^T $ and this Norrell state be $|\psi_N\ket = \frac{1}{\sqrt{6}} (-1,2,-1)^T$.  Let us consider the strange state (Norrell state) $|\psi_{S}\ket (|\psi_{N}\ket)$ mixed with the maximally mixed state to have the 
family of states $\rho_{S} (\rho_{N}) =  (1-p)|\psi_{S} \ket\bra \psi_{S}| + \frac{p}{3} \mathbb{I} \quad ( (1-p)|\psi_{N} 
\ket\bra \psi_{N}| + \frac{p}{3} \mathbb{I})$. With the explicit expression for sum negativity given previously, it is easy to check that the sum negativity for the noisy strange state is given by 

 \begin{equation}
M_{SN} [\rho_S]=\left\lbrace
\begin{array}{@{}ll@{}}
\frac{2}{9} (3 - 4p), & \text{if}\ 0 \leq p \leq \frac{3}{4} \\
0,  & \text{if}\  \frac{3}{4} \leq p \leq 1  
\end{array}\right\rbrace
\end{equation} 
while the corresponding sum negativity for the noisy Norrell state is given by 
 \begin{equation}
M_{SN} [\rho_N]=\left\lbrace
\begin{array}{@{}ll@{}}
\frac{2}{9} (3 - 5p), & \text{if}\ 0 \leq p \leq \frac{3}{5} \\
0,  & \text{if}\  \frac{3}{5} \leq p \leq 1 
\end{array}\right\rbrace
\end{equation} 

\noindent Therefore, we see that the strange state remains more robust against admixture with white noise than the Norrell state.\qed

Now, let us consider an example of a purely coherent noise, i.e. admixture of a maximally magical pure state with a maximally coherent state $|c\ket = \frac{1}{\sqrt{3}}(|0\rangle - |1\rangle + |2\rangle)$. 

\emph{Proposition - The Norrell state above is more robust under the admixture of aforementioned coherent noise than the  strange state above.}

\emph{Proof-} Proceeding similarly as before, the expression for sum negativity of the noisy strange state is now given by \begin{equation}
M_{SN} [\rho_S]=\left\lbrace
\begin{array}{@{}ll@{}}
\frac{2}{9} (3 - 2p), & \text{if}\ 0 \leq p \leq \frac{3}{5} \\
\frac{1}{9} (3  + p), & \text{if}\  \frac{3}{5} \leq p \leq 1  
\end{array}\right\rbrace,
\end{equation} 

while the corresponding expression for sum negativity of the noisy Norrell state is given by 
\begin{equation}
M_{SN} [\rho_N] = \frac{2}{9} (3 - p).
\end{equation}

Thus, throughout the range of the noise parameter $p$, the noisy Norrell state contains more 'magic' than the corresponding noisy strange state, which is demonstrated in Fig. \ref{cohnoise}.
\subsection{Relation betweeen quantum coherence, quantum entanglement, and 'magic'}

\begin{figure}
\centering
\subfigure[Magic (quantified by sum negativity) vs $l_1$-norm coherence for randomly chosen qutrit pure (deep blue) and mixed (light blue) states. The red line corresponds to the bound conjectured in \eqref{coh_conj}.]{
    \includegraphics[width=0.4\textwidth, keepaspectratio]{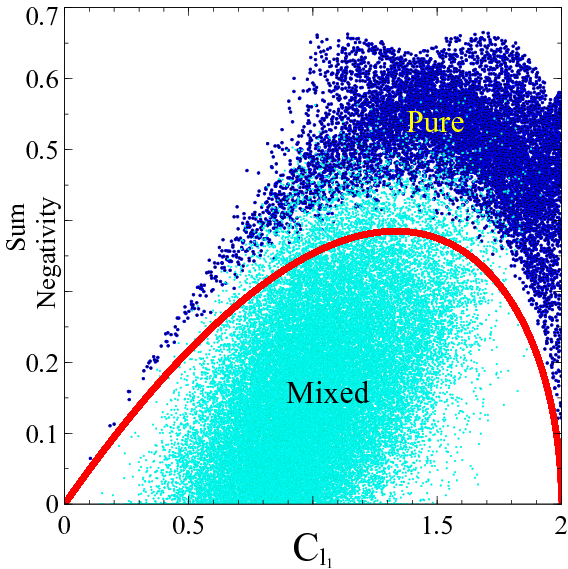}
    \label{cohbound}
}
\subfigure[Magic (quantified by sum negativity) of the reduced qutrit system vs negativity measure of entanglement for randomly chosen qutrit-qubit pure (orange) and mixed (green) states. The blue line corresponds to the bound conjectured in \eqref{ent_conj}. ]{
    \includegraphics[width=0.4\textwidth, keepaspectratio]{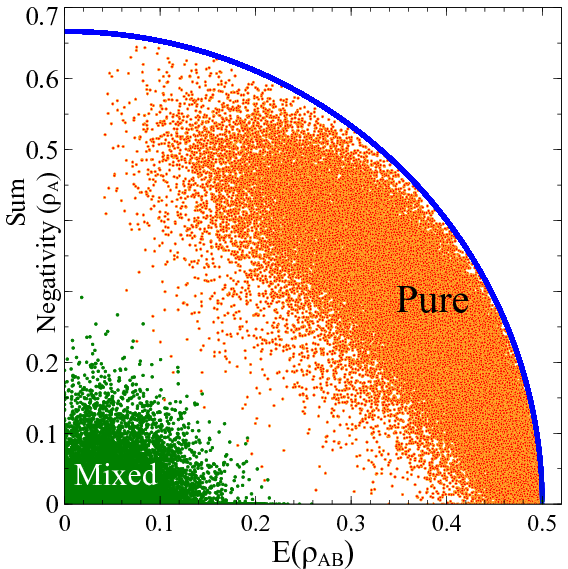}       
    \label{entbound}
}
\caption{(Color online) Interplay between quantum coherence, 'magic' and entanglement in a qutrit (left) and a qutrit-qubit system (right).}
\label{bounds}
\end{figure}

Continuing with our theme of attempting to unearth the relation of coherence and 'magic' in quantum systems, it is a natural question to ask whether we can find a bound for the quantity of 'magic' in terms of coherence in the qutrit scenario. One bound is quite obvious. Every incoherent state lies within the stabilizer polytope, therefore it is easy to see that any quantum state, pure or mixed, is at least as close to a stabilizer state as to an incoherent state. Thus, the magic of a quantum state is upper bounded by the amount of coherence in the system. However, for qutrit pure states, numerical simulation in Fig. \ref{cohbound} leads us to conjecture the following inequality, which gives a reverse, i.e., lower bound to the magic in terms of quantum coherence. 

\emph{Proposition - The following condition on quantum coherence, quantified via the $l_1$-norm, and 'magic', quantified by the sum negativity, holds for 
qutrit pure states} \beq M_{SN}[|\psi\ket] \geq \frac{C_{l_{1}}[|\psi\ket]}{2} \sqrt{1- \frac{C_{l_{1}}[|\psi\ket]}{2}} \label{coh_conj}\eeq

It has already been shown that the presence of entanglement in a bipartite state adversely affects the coherence \cite{mcms,hall} as well as contextuality \cite{cont_vs_magic}  in the reduced state. Since 'magic' as a resource in quantum computation has ultimately been ascribed to the contextual nature of quantum mechanics \cite{contextuality_supplies_magic}, it is important to quantify the corresponding trade off for entanglement in the joint system and magic in the reduced system. The simplest case is that of a qutrit qubit joint system.  In this situation, we conjecture the following trade off relation between bipartite entanglement, quantified by the negativity, of a qutrit qubit joint system $AB$, and that of 'magic', quantified by sum negativity, in the reduced qutrit system $A$.

\emph{Proposition - The negativity of entanglement $E_{AB}$ and the sum negativity $SN_{A}$ satisfies the following trade off relation } 
\beq 16 E_{AB}^{2} + 9 M_{SN_{A}}^{2} < 4 \label{ent_conj} \eeq

Although an analytical proof is lacking, the numerical result furnished in Fig. \ref{entbound} strongly suggests that the proposition above is true and indeed, almost tight for pure $3\times2$ states.

\section{Conclusion}

Unification is a common theme in physics. Following the spirit of unification, it is thus a worthwhile effort to bring various nascent resource theories in quantum information theory under one umbrella. In this work, we have indicated the link between the resource theories of coherence and 'magic'. We demonstrated that quantum coherence in a state is ultimately the currency for creation of magic states through incoherent operations. Furthermore, we derived another full coherence monotone from the discrete Wigner function representation of quantum states in discrete phase space. We also proposed several sub-classes of quantum operations as free operations if coherence and 'magic' are to be simultaneously considered as resources. We also investigated the link between coherence and 'magic' in the concrete scenario of a qutrit system. We believe the concepts and results in this paper should spur more detailed investigations into the link between these two different resource theories. More generally, the method utilized to prove \eqref{bhou} is almost identical to the corresponding proof in Ref. \cite{uttament}, which indicates that a more general result along these lines can be proved in other situations as well. Our work suggests that various resources useful for quantum technology are ultimately the manifestation of the superposition principle in the quantum world, quantified through the resource theory of coherence. We also welcome work expounding on the link between the present stabilizer formalism and the studies on non-Gaussianity for CV systems in recent literature \cite{adesso_gau, takagi, albarelli}

 \section{Acknowledgement}
CM thanks Department of Atomic Energy, Govt of India for financial support through fellowship. We also acknowledge Uttam Singh, Samyadeb Bhattacharya, and Victor Veitch for useful inputs. We thank Francesco Albarelli for pointing out an error in an earlier version of the manuscript and the anonymous referee for suggesting improvements.

\vspace{0.6 in} 
\section*{References}
\bibliographystyle{iopart-num}
\bibliography{magic_ref}

\end{document}